\documentclass{article}

\usepackage{spconf,amsmath,graphicx,hyperref}

\title{MIDI-LLaMA: An Instruction-Following Multimodal LLM for Symbolic Music Understanding
\vspace{-2mm}}

\name{Meng Yang$^{1}$ \qquad
      Jon McCormack$^{1}$ \qquad
      Maria Teresa Llano$^{2}$ \qquad
      Wanchao Su$^{1}$ \qquad
      Chao Lei$^{3}$
      \vspace{-2mm}}

\address{$^{1}$ SensiLab, Monash University, Australia \\
         $^{2}$ University of Sussex, Brighton, United Kingdom \\
         $^{3}$ School of Computing and Information Systems, The University of Melbourne, Australia
         \vspace{-3mm}}

\begin{document}
%
\maketitle
\vspace{-4mm}
\begin{abstract}

Recent advances in multimodal large language models (\allowbreak MLLM) for audio music have demonstrated strong capabilities in music understanding, yet symbolic music, a fundamental representation of musical structure, remains unexplored. In this work, we introduce MIDI-LLaMA, the first instruction-following MLLM for symbolic music understanding. Our approach aligns the MIDI encoder MusicBERT and Llama-3-8B via a two-stage pipeline comprising feature alignment and instruction tuning. To support training, we design a scalable annotation pipeline that annotates GiantMIDI-Piano with fine-grained metadata, resulting in a MIDI–text dataset. Compared with the baseline trained on converting MIDI into ABC notation under the same instruction-tuning procedure, MIDI-LLaMA substantially outperforms in captioning and semantic alignment in question answering. Human evaluation further confirms the advantages of MIDI-LLaMA in music understanding, emotion recognition, creativity, and overall preference. These findings demonstrate that incorporating symbolic music into large language models enhances their capacity for musical understanding.
\end{abstract}
\vspace{-2mm}
\section{Introduction}
\label{sec:intro}

Music Understanding is a core interest of the Music Information Retrieval (MIR) community. Classical MIR pipelines map music to predefined tags, such as tempo \cite{tempo1}, key \cite{key1}, and emotion \cite{emotion1}, enabling tasks like similarity search and tagging, but restricting the system to closed-set questions defined by the feature taxonomy. Recently, LLMs have opened new directions in MIR by enabling open-ended interaction beyond fixed tag taxonomies, supporting applications such as conversational score tutoring \cite{chatmusicians}, prompt-based generation \cite{chatmusicians}, automated captioning \cite{llark,musilingo}, and assistive composition \cite{composition}.

Recent advances in multimodal large language models (MLLMs) have shown that instruction tuning enables open-ended cross-modal interaction. In vision–language research, LLaVA \cite{llava} demonstrated that aligning a frozen CLIP \cite{clip} encoder with an LLM via visual instruction tuning can approach GPT-4-level performance on open-domain dialogue and image reasoning. Building on this paradigm, LLark \cite{llark} aligns LLMs with audio encoders to support music-grounded dialogue and improve music understanding, with related music–language alignment approaches also explored \cite{musilingo,music_llama}. However, these efforts remain limited to audio music. In parallel, NOTA \cite{nota} introduces multimodal music-notation understanding for visual LLMs, tackling music scores from a vision-based perspective, but still does not address symbolic music itself as a distinct modality. Symbolic music, however, provides clearer structure, lower noise, and greater editability than audio \cite{musicbert}, and has long been important in MIR. Progress has been constrained by the absence of large-scale symbolic music–text datasets. As a result, researchers often approximate symbolic music, such as MIDI, by converting it into ABC notation, a lightweight, text-based representation, for use in text-only models \cite{ABC_Notation1, chatmusicians}. While convenient, this conversion discards rhythmic and polyphonic details, limiting the depth of musical understanding.

To bridge this gap, we propose MIDI-LLaMA, the first symbolic music (MIDI)-text instruction-following multimodal Language Model. We adopt MusicBERT as the MIDI music encoder and integrate it with the open-source Llama-3-8B language model, completing end-to-end alignment through a two-stage pipeline similar to LLaVA and LLark. For training data, we employ GPT-4o-based textual metadata mining and manual verification to annotate the GiantMIDI-Piano dataset \cite{GiantMIDI} with genre, style, perceived emotion, and expressive intent, creating the first symbolic music–text dataset focused on classical piano music.

We evaluate MIDI-LLaMA on music understanding question answering and music captioning tasks using BLEU \cite{BLEU}, METEOR \cite{METEOR}, ROUGE-L \cite{ROUGE}, and BERTScore \cite{BERTScore}. Since no prior symbolic-music multimodal baseline exists, we construct a text-only baseline trained on the same data and instruction-tuning procedure with ABC notation as input. MIDI-LLaMA consistently outperforms this baseline across both tasks and most metrics, demonstrating that integrating symbolic music encodings with LLMs via instruction tuning substantially improves symbolic-music understanding. Human evaluation further shows a clear preference for MIDI-LLaMA’s captions, which are rated as more accurate in music understanding and capturing musical emotion.

Our main contributions are as follows: (1) We present the first symbolic music (MIDI)-text instruction-following multimodal LLM. While building on existing components, this is the first study to adapt such an architecture to the symbolic-music modality. (2) We develop an automated pipeline combining GPT-assisted metadata mining with manual verification to generate paired music–text data. The pipeline produces high-quality annotations for over 9,800 known pieces with human validation showing 89\% acceptance for categorical tags and 93\% for descriptive annotations, helping mitigate the scarcity of symbolic music–text data.

\vspace{-2mm}
\section{Dataset}
\label{sec:format}

\subsection{Automatic Dataset Annotation Pipeline}

The scarcity of large-scale symbolic music–text datasets has significantly constrained multimodal research in this domain. Manual annotation, while reliable, quickly reaches its limits in scale due to its prohibitive cost and time requirements. To address this, we design an LLM-based automatic annotation pipeline for music annotation.

Among existing studies, the only relevant prior dataset we are aware of is MidiCaps \cite{midiCaps}, which pairs MIDI with textual tags and generates captions from them. However, all tags in MidiCaps are derived from MIR algorithms, which can be limited in accuracy, and labels tend to be general. In contrast, our method begins from music-related contextual information rather than algorithmic predictions. Specifically, we query online sources using the title and composer as keywords, crawl the music-related introduction from reputable classical music websites, and provide the retrieved text to GPT-4o. The GPT-4o is prompted to extract genre, style, composition background and expressive intent, and to infer perceived emotion from the context. To improve reliability, the prompt includes a \textit{Not Enough Information} option when contextual resources are insufficient, preventing the model from fabricating content. We further set the temperature to 0 to enforce deterministic outputs and reduce hallucinations.

We apply this pipeline to annotate GiantMIDI-Piano \cite{GiantMIDI}, a classical piano MIDI dataset containing 10,855 MIDI files from 2,786 composers. Following the annotation pipeline, a total of 9,803 musical pieces were assigned valid tags. Our annotation covers a fine-grained set of annotations, producing both structured labels and free-form descriptions that support detailed genre-specific music understanding.

\vspace{-2mm}
\subsection{Annotation Quality Evaluation}
To assess the reliability of the automatically generated annotations, we performed a stratified sampling of 100 annotated works across different composers and periods. Two domain experts independently reviewed the automatically extracted labels against the retrieved web sources, marking them as unacceptable if they contained clear factual errors and as acceptable if broadly consistent with the source information. This evaluation shows that short categorical tags achieved an acceptance rate of 89\%, while free-form descriptive fields (compositional background and expressive intent) reached an acceptance rate of 93\%. Since emotion annotation is relatively subjective, our previous work introduced a human-validated framework showing that LLM-based emotion annotation performs comparably to expert annotation \cite{data}. Together, these findings indicate that the proposed pipeline produces annotations of sufficient reliability for downstream multimodal training and evaluation.
\vspace{-2mm}
\subsection{Instruction-Tuning Data Generation}
\label{instruction_tuning}
To construct instruction-tuning data, we first extracted basic musical features from each MIDI file using music21 \cite{music21}, including tempo, key, and time signature. These features were incorporated into the tags as supplementary musical information, aiding the model’s understanding of low-level musical structure and serving as evidence for reasoning about higher-level features such as style and emotion. We then employed GPT-4o to generate natural language question-answer (Q\&A) pairs from the annotated tags. For each tag (e.g., genre, style, background, expressive intent, perceived emotion), GPT-4o was prompted to formulate corresponding questions and provide an answer grounded in the given tag. This process produced multiple Q\&A pairs per piece, covering both factual and interpretive aspects of music understanding.

Since many pieces in GiantMIDI-Piano are of substantial duration, we segmented three non-overlapping 20-second clips from each piece to expand data volume and enhance representational diversity. To maximize coverage, clips were sampled from different positions within the piece (e.g., beginning, middle, and later sections). Segmentation boundaries were aligned with bar lines when feasible, so that clips would contain complete musical measures. In total, the procedure produced 29,409 clips and approximately 2.3 million Q\&A pairs, forming a large-scale corpus for symbolic music–text multimodal learning.

\vspace{-3mm}
\section{Method}

\begin{figure*}[ht!]
  \centering
  \includegraphics[width=\textwidth]{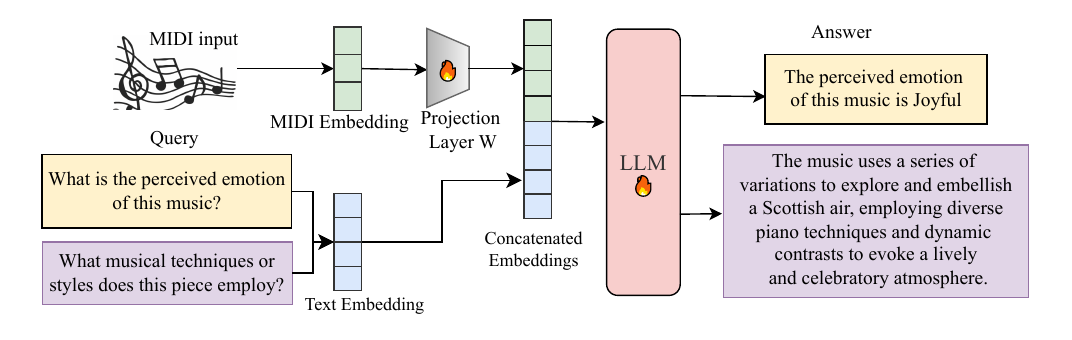}
  \caption{Overall architecture of MIDI-LLaMA: given a MIDI input, the model can answer music-related questions. The real examples are shown in the graph. Components annotated with the flame symbol are trainable modules.}
  \label{fig:wide}
  \vspace{-4mm}
\end{figure*}

\subsection{Model Architecture}
A key challenge in multimodal large language models (MLLMs) is to bridge modality-specific encoders with a language backbone. BLIP-2 \cite{BLIP} introduced a lightweight Q-Former to query frozen vision encoders before injecting features into an LLM. LLaVA \cite{llava} and MiniGPT-4 \cite{minigpt} simplified this approach by using a linear projection layer to map CLIP embeddings directly into the text embedding space, showing that small adapters combined with instruction tuning can already achieve competitive performance. In the audio domain, LLark \cite{llark} applies this two-stage recipe, demonstrating strong improvements in audio music understanding. 

We extend the LLaVA/LLark architecture to the symbolic music (MIDI) modality. The MIDI-LLaMA model consists of a frozen music encoder and a frozen language model, connected via a trainable projection layer that maps music embeddings into the LLM’s text embedding space. This projection produces ``music tokens'' which are concatenated with the textual embeddings, enabling the LLM to jointly process symbolic music and language. For the language backbone, we adopt Llama-3-8B, which balances performance with efficiency and remains feasible to fine-tune under moderate computation budgets. For the symbolic music encoder, we employ MusicBERT\cite{musicbert}, which was pretrained on millions of symbolic music pieces using the OctupleMIDI representation and bar-level masking strategies specifically designed for musical structure. MusicBERT achieves strong baselines on melody completion, accompaniment suggestion, and genre/style classification, suggesting that its embeddings capture core musical features such as rhythm, phrasing, pitch and harmony, making it a suitable encoder for multimodal alignment.

Each MIDI file is tokenized into OctupleMIDI events, which encode eight aspects of a note event (pitch, duration, velocity, bar, position, tempo, time signature, instrument). MusicBERT processes these sequences into token-level hidden states. We then apply temporal mean pooling to obtain an $M$-dimensional clip-level embedding. A trainable projection layer maps this embedding to the hidden size $T$ of Llama-3-8B, producing prefix ``musical tokens'' concatenated with the text embedding,  enabling the LLM to incorporate symbolic music information into its generation process.
\vspace{-2mm}
\subsection{Training Procedure}
Training proceeds in two stages. In the alignment stage, both encoders are kept frozen and only the projection layer is optimized on the instruction-tuning data (Section~\ref{instruction_tuning}) with next-token cross-entropy, ensuring that the projected music embeddings are interpretable to the LLM. In the instruction-tuning stage, the music encoder remains frozen while the projection is further updated and the LLM is adapted with LoRA modules (rank = 8) on the same instruction–tuning Q\&A pairs, enabling diverse musical tasks. Both stages employ the AdamW optimizer with a maximum learning rate of 5e-4, a linear warmup ratio of 0.03 followed by cosine decay, a batch size of 16, and are trained on two A6000 GPUs.
\vspace{-2mm}
\section{Evaluation}
\vspace{-2mm}
\subsection{Baseline}
As no prior multimodal large language model has been developed for symbolic music, we constructed a text-only baseline to evaluate the contribution of incorporating symbolic music embeddings. As discussed in Section \ref{sec:intro}, prior LLM-based approaches to symbolic music typically convert it into ABC Notation and treat it purely as text input. Following this practice, we converted all musical data from our instruction-tuning corpus into ABC Notation and trained a baseline model, ABC-LLaMA, using exactly the same instruction-tuning procedure, training set, and hyperparameters as our proposed MIDI-LLaMA. This ensures that the only difference between the two models is the representation of symbolic music: MIDI embeddings from MusicBERT (MIDI-LLaMA) versus Textual transcriptions ABC Notation (ABC-LLaMA), enabling a direct comparison of the benefit of aligning symbolic music embeddings with the language model.

\vspace{-2mm}
\subsection{Task Setting and Evaluation Metrics}
We evaluate the models on two subtasks by partitioning the test set of our instruction-tuning data: question Answering, where the system responds to queries targeting specific aspects of music understanding (e.g., style or emotion), and Music Captioning, where it generates a concise description of the music overall. For both tasks, outputs are assessed using four widely adopted text-generation metrics, BLEU \cite{BLEU}, METEOR \cite{METEOR}, ROUGE-L \cite{ROUGE}, and BERTScore \cite{BERTScore}. Together, these metrics balance lexical accuracy and semantic fidelity, offering a comprehensive assessment of symbolic music understanding and caption quality.
\vspace{-3mm}
\subsection{Quantitative Results}
\begin{table}[t]
\small
\begin{tabular}{c|cccc}
\hline
\multicolumn{1}{l|}{Model} & \multicolumn{1}{l}{B-U↑} & \multicolumn{1}{l}{M-R↑} & \multicolumn{1}{l}{R-L↑} & \multicolumn{1}{l}{BERT-S↑} \\ \hline
\multicolumn{1}{l|}{}      & \multicolumn{4}{c}{Question Answering}                                                                       \\ \hline
LLaMA-3-8B                 & 0.0004                   & 0.0101                   & 0.0113                   & 0.6077                      \\
LLaMA-3-70B                & 0.0032                   & 0.0211                   & 0.0153                   & 0.4408                      \\
ABC-LLaMA                  & \textbf{0.2352}          & \textbf{0.2792}          & 0.5395                   & 0.8529                      \\
MIDI-LLaMA                 & 0.2001                   & 0.2344                   & \textbf{0.5486}          & \textbf{0.9519}             \\ \hline
\multicolumn{1}{l|}{}      & \multicolumn{4}{c}{Music Captioning}                                                                         \\ \hline
LLaMA-3-8B                 & 0.0467                   & 0.1826                   & 0.1412                   & 0.8335                      \\
LLaMA-3-70B                & 0.0519                   & 0.1910                   & 0.1415                   & 0.8409                      \\
ABC-LLaMA                  & 0.1592                   & 0.2919                   & 0.2607                   & 0.8536                      \\
MIDI-LLaMA                 & \textbf{0.2566}          & \textbf{0.3797}          & \textbf{0.4265}          & \textbf{0.9142}             \\ \hline
\end{tabular}
\caption{Quantitative evaluation on question answer and music captioning. BLEU (B-U↑), METEOR (M-R↑), ROUGE-L (R-L↑), and BERTScore (BERT-S↑, without rescaling) are reported. The best values of metrics are made bold.}
\vspace{-4mm}
\label{result}
\end{table}
The results in Table~\ref{result} show clear gains from incorporating symbolic music embeddings. In question answering, MIDI-LLaMA yields superior ROUGE-L and BERTScore, reflecting stronger semantic alignment, while ABC-LLaMA shows higher BLEU and METEOR, suggesting it benefits from surface-level lexical overlap when answers are short. In music captioning, MIDI-LLaMA achieves a large improvement over ABC-LLaMA across all metrics, indicating a better ability to capture holistic musical content and expression. Importantly, the raw zero-shot LLaMA-3-8B and 70B text-only baselines perform very poorly, confirming that general-purpose LLMs cannot handle symbolic music without adaptation. Overall, these results validate that aligning MusicBERT embeddings with an instruction-tuned LLM substantially enhances symbolic music understanding beyond text-only baselines based on ABC Notation.
\vspace{-3mm}
\subsection{Human Evaluation on Music Captioning}
Since music captioning is subjective, standard quantitative metrics alone cannot fully capture the quality of generated music captions. To complement quantitative evaluation, we conducted a human study comparing captions produced by MIDI-LLaMA and ABC-LLaMA. We randomly sampled 100 musical clips and obtained captions from both models. A total of 17 participants each evaluated an average of 30 pairwise comparisons: for each clip, they listened to the audio and were shown two captions in randomized order. They judged which caption better reflected (i) music understanding accuracy, (ii) text fluency, (iii) creativity, and (iv) musical emotion understanding, and then indicated their overall preference. Each clip was independently judged by 5 participants, yielding 500 judgments in total. Preferences were aggregated by majority vote. This setup enables a direct, human-centred subjective assessment of music caption quality.

\begin{figure}
    \centering
    \includegraphics[width=\linewidth]{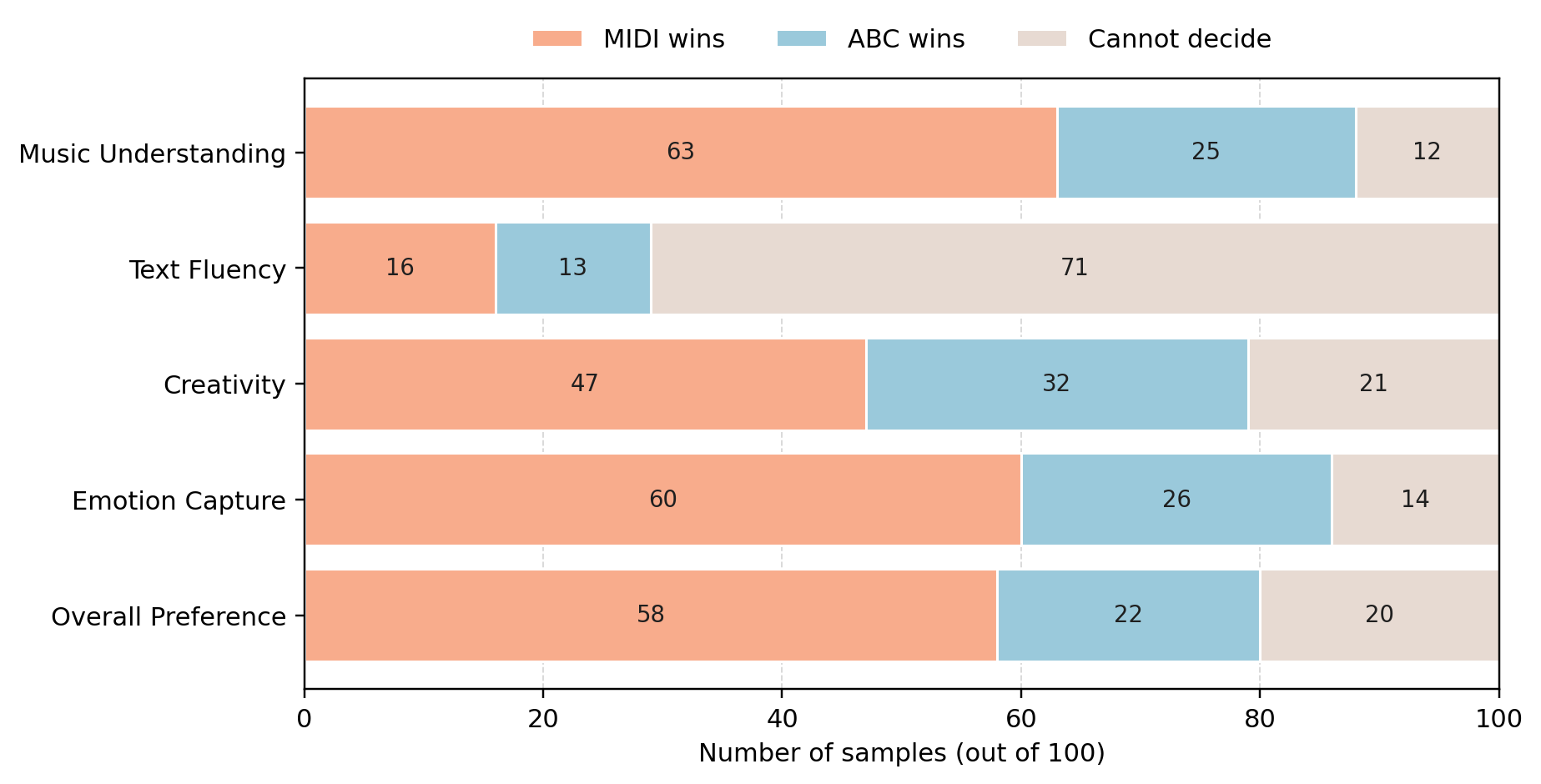}
    \vspace{-6mm}
    \caption{Human comparative evaluation of captions by MIDI-LLaMA and ABC-LLaMA across five dimensions.}
    \label{fig:human_eval}
    \vspace{-3mm}
\end{figure}

The result from human evaluation further corroborates the advantage of incorporating symbolic music embeddings. As shown in Figure~\ref{fig:human_eval}, MIDI-LLaMA was preferred over ABC-LLaMA in the majority of dimensions. In particular, it achieved clear wins on music understanding (63 vs. 25) and emotion recognition (60 vs. 26), indicating that participants judged its captions to more accurately reflect the musical content and affective qualities. For creativity, MIDI-LLaMA was also rated higher (47 vs. 32), suggesting that its descriptions were perceived as more varied and imaginative. Text fluency was the most inconclusive dimension, where a large proportion of raters chose cannot decide (71/100), reflecting that both models produced comparably fluent text. Overall, MIDI-LLaMA was favoured in 58 clips versus 22 for ABC-LLaMA, demonstrating a consistent human preference that aligns with the automatic metrics and highlights the benefit of explicit symbolic embeddings for caption quality.

Our experiments demonstrate that MIDI-LLaMA consistently outperforms the text-only ABC-LLaMA baseline. Automatic metrics show substantial gains in captioning and stronger semantic alignment in question answering, while human evaluation confirms clear advantages in music understanding, emotion recognition, creativity, and overall preference. These results collectively validate the effectiveness of incorporating symbolic music embeddings for instruction-tuned language models. However, our evaluation is restricted to classical piano data. Future work will explore broader repertoires and alternative alignment strategies.

\vspace{-1mm}

\section{Conclusion}
\vspace{-1mm}
We presented MIDI-LLaMA, the first instruction-following multimodal LLM for symbolic music, aligning MusicBERT embeddings with Llama-3-8B through a two-stage training pipeline. To support this, we developed a scalable annotation pipeline and constructed a symbolic music–text dataset from GiantMIDI-Piano. Both automatic metrics and human evaluation demonstrate that MIDI-LLaMA consistently outperforms a carefully designed text-only baseline, confirming the effectiveness of incorporating symbolic music embeddings into LLMs through instruction-tuning. This integration also opens new opportunities for fine-grained symbolic music analysis, interactive composition, and bridging symbolic and audio modalities in future multimodal systems.

\vfill\pagebreak

\bibliographystyle{IEEEbib}
\bibliography{strings,refs}

\begin{thebibliography}{10}

\bibitem{tempo1}
Miguel Alonso, Bertrand David, and Gaël Richard,
\newblock ``Tempo and beat estimation of musical signals,''
\newblock in {\em Proc. 5th Int. Soc. Music Information Retrieval Conf. (ISMIR)}, 10 2004.

\bibitem{key1}
Filip Korzeniowski and Gerhard Widmer,
\newblock ``End-to-end musical key estimation using a convolutional neural network,''
\newblock in {\em 2017 25th European Signal Processing Conference (EUSIPCO)}, 2017, pp. 966--970.

\bibitem{emotion1}
Youngmoo~E Kim, Erik~M Schmidt, Raymond Migneco, et~al.,
\newblock ``Music emotion recognition: A state of the art review,''
\newblock in {\em Proc. 11th Int. Soc. Music Information Retrieval Conf. (ISMIR)}, 2010, vol.~86, pp. 937--952.

\bibitem{chatmusicians}
Ruibin Yuan, Hanfeng Lin, Yi~Wang, et~al.,
\newblock ``{C}hat{M}usician: Understanding and generating music intrinsically with {LLM},''
\newblock in {\em Findings Assoc. Comput. Linguistics (ACL)}, 2024, pp. 6252--6271.

\bibitem{llark}
Josh Gardner, Simon Durand, et~al.,
\newblock ``Llark: a multimodal instruction-following language model for music,''
\newblock in {\em Proceedings of the 41st International Conference on Machine Learning}, 2024, ICML'24.

\bibitem{musilingo}
Zihao Deng, Yinghao Ma, Yudong Liu, et~al.,
\newblock ``{M}usi{L}ingo: Bridging music and text with pre-trained language models for music captioning and query response,''
\newblock in {\em Findings Assoc. Comput. Linguistics (NAACL)}, June 2024, pp. 3643--3655.

\bibitem{composition}
Qixin Deng, Qikai Yang, et~al.,
\newblock ``Composerx: Multi-agent symbolic music composition with llms,'' 2024.

\bibitem{llava}
Haotian Liu, Chunyuan Li, Qingyang Wu, and Yong~Jae Lee,
\newblock ``Visual instruction tuning,''
\newblock in {\em Advances in Neural Information Processing Systems}, 2023, vol.~36, pp. 34892--34916.

\bibitem{clip}
Alec Radford, Jong~Wook Kim, Chris Hallacy, et~al.,
\newblock ``Learning transferable visual models from natural language supervision,''
\newblock in {\em Proc. 38th Int. Conf. Mach. Learn. (ICML}, 2021, pp. 8748--8763.

\bibitem{music_llama}
Shansong Liu, Atin~Sakkeer Hussain, et~al.,
\newblock ``Music understanding llama: Advancing text-to-music generation with question answering and captioning,''
\newblock in {\em Proc. IEEE Int. Conf. Acoust., Speech Signal Process. (ICASSP)}, 2024, pp. 286--290.

\bibitem{nota}
Mingni Tang, Jiajia Li, Lu~Yang, et~al.,
\newblock ``{NOTA}: Multimodal music notation understanding for visual large language model,''
\newblock in {\em Findings Assoc. Comput. Linguistics (NAACL)}, Albuquerque, New Mexico, Apr. 2025, pp. 7160--7173.

\bibitem{musicbert}
Mingliang Zeng, Xu~Tan, Rui Wang, et~al.,
\newblock ``{M}usic{BERT}: Symbolic music understanding with large-scale pre-training,''
\newblock in {\em Findings of the Association for Computational Linguistics: ACL-IJCNLP 2021}, Aug. 2021, pp. 791--800.

\bibitem{ABC_Notation1}
Xingwei Qu, yuelin bai, et~al.,
\newblock ``Mu{PT}: A generative symbolic music pretrained transformer,''
\newblock in {\em Proc. 13th Int. Conf. Learning Representations (ICLR)}, 2025.

\bibitem{GiantMIDI}
Qiuqiang Kong, Bochen Li, Jitong Chen, and Yuxuan Wang,
\newblock ``Giantmidi-piano: A large-scale midi dataset for classical piano music,''
\newblock {\em Transactions of the International Society for Music Information Retrieval}, May 2022.

\bibitem{BLEU}
Kishore Papineni, Salim Roukos, et~al.,
\newblock ``Bleu: a method for automatic evaluation of machine translation,''
\newblock in {\em Proc. 40th Annu. Meet. Assoc. Comput. Linguistics (ACL)}, 2002, p. 311–318.

\bibitem{METEOR}
Alon Lavie and Abhaya Agarwal,
\newblock ``Meteor: an automatic metric for mt evaluation with high levels of correlation with human judgments,''
\newblock in {\em Proceedings of the Second Workshop on Statistical Machine Translation}, 2007, p. 228–231.

\bibitem{ROUGE}
Chin-Yew Lin,
\newblock ``{ROUGE}: A package for automatic evaluation of summaries,''
\newblock in {\em Text Summarization Branches Out}, July 2004, pp. 74--81.

\bibitem{BERTScore}
Tianyi Zhang, Varsha Kishore, et~al.,
\newblock ``Bertscore: Evaluating text generation with {BERT},''
\newblock {\em CoRR}, vol. abs/1904.09675, 2019.

\bibitem{midiCaps}
Jan Melechovsky, Abhinaba Roy, and Dorien Herremans,
\newblock ``Midicaps: A large-scale midi dataset with text captions,''
\newblock in {\em Proc. 25th Int. Soc. Music Inf. Retrieval Conf. (ISMIR)}, Nov 2024.

\bibitem{data}
Meng Yang, Jon McCormack, et~al.,
\newblock ``Exploring the feasibility of llms for automated music emotion annotation,'' 2025.

\bibitem{music21}
Michael~Scott Cuthbert and Christopher Ariza,
\newblock ``Music21: A toolkit for computer-aided musicology and symbolic music data.,''
\newblock in {\em Proc. 11th Int. Soc. Music Information Retrieval Conf. (ISMIR)}, 2010.

\bibitem{BLIP}
Junnan Li, Dongxu Li, et~al.,
\newblock ``Blip-2: bootstrapping language-image pre-training with frozen image encoders and large language models,''
\newblock in {\em Proc. 40th Int. Conf. Mach. Learn. (ICML)}, 2023, ICML'23.

\bibitem{minigpt}
Deyao Zhu, Jun Chen, et~al.,
\newblock ``Mini{GPT}-4: Enhancing vision-language understanding with advanced large language models,''
\newblock in {\em Proc. 12th Int. Conf. Learning Representations (ICLR)}, 2024.

\end{thebibliography}
\small

\end{document}